\documentclass[aps,pra,amsmath,amssymb,superscriptaddress,preprint]{revtex4-1}
\usepackage[english]{babel}
\usepackage{bbm}
\usepackage{amsmath}
\usepackage{verbatim}
\usepackage{graphicx}
\usepackage{times}
\usepackage{amssymb}
\usepackage{epsfig}
\usepackage{graphicx}
\usepackage{bm}
\usepackage{times}
\usepackage{txfonts}
\usepackage{color}
\usepackage[normalem]{ulem}

\newcommand{\la}{\langle}
\newcommand{\ra}{\rangle}
\newcommand{\da}{\dagger}

\newcommand{\SN}{\mathrm{SN}}
\newcommand{\diff}{\mathrm{d}}
\newcommand{\vk}{\boldsymbol{k}}
\renewcommand{\vr}{\boldsymbol{r}}
\newcommand{\vq}{\boldsymbol{q}}
\newcommand{\vs}{\boldsymbol{s}}

\newcommand{\Op}[1]{\mathsf #1}

\newcommand{\oA}{ \Op{A} }
\newcommand{\oM}{ \Op{M} }
\newcommand{\oB}{ \Op{B} }

\newcommand{\oH}{ \Op{H} }
\newcommand{\ovr}{ \text{\sf \textbf{r}}}
\newcommand{\ovp}{ \text{\sf \textbf{p}}}

\newcommand{\erw}[1]{ {\cal E} \left[ #1 \right]}

\begin{document}

\title{Stochastic extensions of the regularized Schr\"{o}dinger-Newton equation}

\author{Stefan Nimmrichter}

\affiliation{Faculty of Physics, University of Duisburg-Essen, Lotharstra\ss{}e 1, 47048 Duisburg, Germany}

\author{Klaus Hornberger}

\affiliation{Faculty of Physics, University of Duisburg-Essen, Lotharstra\ss{}e 1, 47048 Duisburg, Germany}

\date\today

\begin{abstract}
We show that the Schr\"{o}dinger-Newton equation, which describes the nonlinear time evolution of self-gravitating quantum matter, can be made compatible with the no-signaling requirement by elevating it to a stochastic differential equation.
In the deterministic form of the equation, as studied so far, the nonlinearity would lead to diverging energy corrections for localized wave packets and would create observable correlations admitting faster-than-light communication. By regularizing the divergencies and adding specific random jumps or a specific Brownian noise process, the effect of the nonlinearity vanishes in the stochastic average and gives rise to a linear and Galilean invariant evolution of the density operator.
\end{abstract}

\maketitle

\section{Introduction}

The Schr\"{o}dinger-Newton (SN) equation has gained growing attention as a possibility both to explain the absence of quantum superpositions at the macro-scale and to reconcile nonrelativistic quantum mechanics with classical Newtonian gravity \cite{Ruffini1969,Diosi1984,Carlip2008,Giulini2011,Giulini2012}. 
According to this equation, the wave function $\psi(\vr)$ of a test mass $m$ creates its own potential energy through gravitational interaction with its ``mass density'' $m|\psi(\vr)|^2$, thereby contributing a nonlinear term to the Schr\"{o}dinger equation. The validity of this approach is, however, still under debate. 

For example, the SN equation is often assumed to be the low-energy manifestation of relativistic gravity in the dynamics of nonrelativistic quantum matter. 
Indeed, the SN equation can be obtained from a semiclassical mean-field approximation of gravitating relativistic field dynamics \cite{Anastopoulos2014,Giulini2014}.
But this only holds in the many-particle limit, where each particle interacts with the gravitational field of its own and all the other masses in the system. Hence it is not clear whether this description makes sense in the case of a single particle. 
In addition, the self-gravitational potential diverges as the particle's wave function covers an increasingly broad range of momenta, approaching, say, a position eigenstate. As a nonrelativistic low-energy approximation, the SN equation should certainly not be applied to such limiting cases. 

Another conceptual issue is related to the nonlinear deterministic nature of the SN equation. It is well known \cite{Gisin1989,Bassi2003} that such nonlinearities may facilitate superluminal communication via entangled states.

Moreover, a thorough analysis of whether the SN equation can help explaining the quantum-classical transition and turn delocalized into localized wave packets has barely begun \cite{Carlip2008,Giulini2011,Bahrami2014a}. Gravitationally-induced or spontaneous collapse models \cite{Karolyhazy1966,Ghirardi1986,Diosi1987,Ghirardi1990b,Penrose1996,Bassi2012} are well-studied alternatives to the SN equation when it comes to the objective reduction of quantum superposition states. It was pointed out within this context that any nonlinear addition to the Schr\"{o}dinger equation, such as the SN potential, must be complemented by an appropriate stochastic term in order to meet the no-signalling condition \cite{Bassi2003,Adler2004,Bahrami2014a}. This should result in a linear master equation for the statistically averaged density operator describing a gradual decay of quantum coherence similar to the predictions of standard decoherence theory \cite{Vacchini2007b,Vacchini2009}. 
On the level of the wave function the previously deterministic time evolution is then affected by discrete jumps or continuous noise. 

Here, we present two simple stochastic extensions of the SN equation; one with discrete jumps determined by a Poissonian random process, and one with continuous white noise following a Wiener process. Both comply with the no-signalling constraint and result in a linear, Markovian, and Galilean-covariant master equation---thus providing a bridge between the SN equation and objective collapse models. 
For this, it will be necessary to operate with a regularized version of the SN equation, where a high-energy cutoff in the gravitational potential implements the above mentioned limitation of the SN equation to low-energy wave functions. Divergent energies, which would lead to conceptual problems in the statistically averaged time evolution, are thus avoided from the start.
We first restrict to the simple and instructive single-particle case, and postpone the treatment of the general $N$-body problem to Sect.~\ref{sec:manybody}.

\section{Regularized Schr\"{o}dinger-Newton equation}

The SN equation,
\begin{equation}
 i\hbar \partial_t \psi(\vr) = \left[ -\frac{\hbar^2}{2m} \Delta + V(\vr)\right] \psi(\vr) + V^\SN_{\psi} (\vr) \psi(\vr), \label{eq:SNE}
\end{equation} 
was proposed as a nonlinear modification of the standard Schr\"{o}dinger equation to describe the influence of classical gravity on the quantum motion of matter \cite{Rosenfeld1963,Diosi1984,Carlip2008,Giulini2012}. Given the wave function $\psi(\vr)$ for a single particle of mass $m$, the nonlinear modification describes the gravitational self-interaction potential of the particle with its own mass probability density $\mu_{\psi}(\vr_0) = m\left| \psi(\vr_0) \right|^2$,
\begin{equation}
 V^\SN_{\psi} (\vr) 
 = -G m \int \diff^3 r_0 \frac{\mu_{\psi}(\vr_0)}{\left| \vr - \vr_0 \right|}.\label{eq:VSN}
\end{equation} 
The SN potential is unbounded in the sense that it diverges if applied to eigenstates of the position operator. Its contribution to the potential energy exceeds all bounds as the particle's wave function gets increasingly localized in space, i.e.~delocalized in momentum. 
For a Gaussian wave packet of spatial width $\sigma$, the expectation value of the SN term scales like $G m^2/\sigma$, whereas the average kinetic energy scales in proportion to $\hbar^2/m\sigma^2$.
Clearly, the SN equation (\ref{eq:SNE}) ceases to be applicable as soon as relativistic energy scales are reached, e.g.~for nucleons at $\sigma\sim 1\,$fm. Yet, at this point the average kinetic energy of the point particle exceeds the SN energy correction by orders of magnitude. 

A standard practice to avoid divergence beyond the nonrelativistic low-energy domain 
is to introduce an effective high-energy cutoff by replacing the $\delta$-peaked mass density in the gravitational potential with a regular function,
\begin{equation}
 \mu (\vr_0) = m \la \psi |\delta(\ovr-\vr_0)|\psi\ra \to m \la \psi |\tilde{g}(|\ovr-\vr_0|)|\psi\ra, 
\end{equation} 
where $\int\diff^3 r \, \tilde{g}(\vr) = 1$ and $\ovr$ the position operator. 
The necessity of this regularization will be discussed below, after introducing stochastic extensions of the SN equation; the unregularized case is restored by setting $g(k)=1$.

With the help of the Fourier transforms $\int\diff^3 r \, e^{i\vk\cdot\vr}/r = 4\pi/k^2$ and $g(k) = \int \diff^3 r \, \tilde{g}(r) e^{i\vk\cdot\vr}$, the regularized SN modification can be expressed in terms of the nonlinear operator
\begin{equation}
 \oH^\SN_{\psi} = V^\SN_{\psi} (\ovr) = - \int\diff^3 k \, \frac{Gm^2}{2\pi^2 k^2} g(k) \la \psi | e^{i\vk\cdot\ovr} |\psi\ra  e^{-i\vk\cdot\ovr}. \label{eq:HSN}
\end{equation} 
It complements a given system Hamiltonian $\oH = \ovp^2/2m + V(\ovr)$. 
In terms of the defined operators, the nonlinear SN equation (\ref{eq:SNE}) reads as $i\hbar \partial_t |\psi\ra = \left( \oH + \oH^\SN_{\psi} \right)|\psi\ra$.

Numerical studies indicate that the (unregularized) SN term prevents the dispersion of wave packets for sufficiently macroscopic masses \cite{Giulini2011}. However, it is not yet clear whether this term will generally turn delocalized wave functions into sufficiently localized classical states, a feature required to explain the quantum-classical transition at the macroscale \cite{Bahrami2014a}. 

Moreover, such a nonlinear equation would in principle allow for superluminal information transfer \cite{Gisin1989,Bassi2003}: 
One could construct an entangled bipartite state, where the time evolution of the reduced state on one side would depend on the choice of measurement basis on the other, arbitrarily distant, side. This problem can be alleviated by adding an appropriate stochastic term to the SN equation which restores the linear time evolution of the statistically averaged density operator  \cite{Bassi2003,Bahrami2014a}. 

\section{Stochastic extensions} \label{sec:extensions}

Let us now present two stochastic extensions of particularly compact form, which will give rise to the same master equation (\ref{eq:ME}). They are determined by the same nonlinear operator 
\begin{equation}
 \oA_{\psi} (\vk) = e^{-i\vk\cdot\ovr} + i \la \psi | e^{-i\vk\cdot\ovr} |\psi\ra. \label{eq:A}
\end{equation} 
The first is a piecewise deterministic extension,
\begin{equation}
 |\diff \psi \ra = -\frac{i}{\hbar} \left( \oH + \oH^\SN_{\psi} \right) |\psi\ra \diff t + \int\diff^3 k \left[ \frac{\oA_{\psi} (\vk)}{\left\| \oA_{\psi} (\vk) |\psi\ra \right\|} - 1 \right] |\psi\ra \diff N_{\vk}, \label{eq:SSE_PDP}
\end{equation} 
with $\diff N_{\vk}$ a family of Poissonian stochastic increments characterized by Eq.~(\ref{eq:dN}) below.
Alternatively, one can consider a diffusive extension (using It\^{o} calculus), 
\begin{eqnarray}
 |\diff \psi \ra &=& -\frac{i}{\hbar} \left( \oH + \oH^\SN_{\psi} \right) |\psi\ra \diff t + \int\diff^3 k \, \oA_{\psi} (\vk) |\psi\ra \diff W_{\vk} \nonumber \\
 && - \frac{1}{2} \int\diff^3 k \frac{Gm^2}{2\pi^2 \hbar k^2} g(k) \oA^\da_{\psi} (\vk) \oA_{\psi} (\vk) |\psi\ra \diff t, \label{eq:SSE_W}
 \end{eqnarray} 
with $\diff W_{\vk}$ a family of Wiener stochastic increments, see Eq.~(\ref{eq:dW}) below. 

\subsection{Piecewise deterministic extension}

The first stochastic extension (\ref{eq:SSE_PDP}) of the regularized SN equation describes discrete quantum jumps interrupting the unitary time evolution of the state vector $|\psi\ra$, as governed by a given system Hamiltonian $\oH$ plus the nonlinear SN term (\ref{eq:HSN}). 
A multivariate Poisson process $N_{\vk} (t)$ \cite{Breuer2002,Wiseman2010} shall decide which jump event (labeled by $\vk$) occurs at what time; a jump associated with the momentum $\hbar\vk$ corresponds to the nonlinear, norm-preserving state transformation $|\psi\ra \to \oA_{\psi} (\vk) |\psi\ra / \left\| \oA_{\psi} (\vk) |\psi\ra \right\|$. 

The Poissonian increments $\diff N_{\vk}$ determine whether or not a jump occurs within the time interval $[t,t+\diff t)$. They are statistically independent, $\erw{\diff N_{\vk} \diff N_{\vq}} = \erw{\diff N_{\vk}} \delta (\vk-\vq)$.
In order for the SN nonlinearity to vanish in the statistical average, the increments must have the state-dependent expectation values
\begin{eqnarray}
  \erw{\diff N_{\vk}} &=& \frac{Gm^2}{2\pi^2 \hbar k^2} g(k) \left\|\oA_{\psi} (\vk) |\psi\ra \right\|^2 \diff t \nonumber \\
 &=& \frac{Gm^2}{2\pi^2 \hbar k^2} g(k) \left( 1 + \left| \la \psi | e^{-i\vk\cdot\ovr} |\psi\ra \right|^2 \right) \diff t \label{eq:dN}.
\end{eqnarray} 
This can be easily confirmed by computing the expectation value $\erw{\diff\left( |\psi\ra \la \psi| \right)}$ and dropping all terms of higher order than $\diff t$. Note that positivity requires $g(k)\geq 0$, and a finite overall jump rate is guaranteed when $\int_0^{\infty}\diff k\, g(k) <\infty$.

\subsection{Diffusive extension}

The second stochastic SN equation (\ref{eq:SSE_W}) describes a Brownian-type diffusion of the state vector in Hilbert space, formulated in It\^{o} calculus \cite{Breuer2002,Wiseman2010}. It is governed by a (state-independent) multivariate complex white noise process $W_{\vk} (t)$, with centered Wiener increments $\diff W_{\vk}$ describing independent random variables, $\erw{\diff W_{\vk}}=0$ and 
\begin{equation}
 \erw{\diff W^{*}_{\vk} \diff W_{\vq}} = \delta(\vk-\vq) \frac{Gm^2}{2\pi^2 \hbar k^2} g(k) \diff t. \label{eq:dW}
\end{equation} 
The noise effect on the state vector is again described by the operator (\ref{eq:A}). Since this transformation does not preserve the norm, the second line is required in the stochastic SN equation (\ref{eq:SSE_W}). It does not contribute to the coherent part of the time evolution, which is again determined by $\oH + \oH_{\psi}^\SN$. As before, the regularizing function $g(k)$ must be positive and integrable.

Although the diffusive extension (\ref{eq:SSE_W}) leads to very different individual quantum trajectories of the state vector,
it is straightforward to show, using the well-known rules of It\^{o} calculus \cite{Breuer2002,Wiseman2010}, that the statistical average assumes the same linear time evolution as in the piecewise deterministic case.

\subsection{Effective classicalization in the ensemble average} 

Both presented stochastic extensions of the regularized SN equation give rise to the same time evolution of the statistically averaged state operator $\rho = \erw{|\psi\ra\la\psi|}$ for the motion of a single particle. It is described by the Lindblad-type master equation,
\begin{equation}
 \partial_t \rho = -\frac{i}{\hbar} \left[ \oH,\rho \right] + \int\diff^3 k \, \frac{Gm^2 g(k)}{2\pi^2 \hbar k^2} \left[ e^{-i\vk\cdot\ovr} \rho e^{i\vk\cdot\ovr} -\rho \right]. \label{eq:ME}
\end{equation} 
This result connects the SN equation with standard collapse models, such as the Di\'{o}si-Penrose (DP) model of gravitational collapse \cite{Diosi1987,Diosi1989,Penrose1996} or the theory of continuous spontaneous localization (CSL) \cite{Ghirardi1990b,Bassi2003}, since these can all be brought to the form (\ref{eq:ME}) by an appropriate choice of the positive function $g(k)$ of finite width. 

The DP model, for instance, always assumes a finite extension of the particle's mass \cite{Diosi1987,Bahrami2014b} from the start to avoid divergencies. Di\'{o}si's original master equation reads as
\begin{equation}
 \partial_t \rho = -\frac{i}{\hbar} \left[ \oH,\rho \right] - \frac{G}{2\hbar} \int \frac{\diff^3 s_1 \diff^3 s_2}{|\vs_1 - \vs_2|} \left[ \varrho \left( \vs_1 - \ovr \right), \left[ \varrho \left( \vs_2 - \ovr \right), \rho \right] \right], \label{eq:ME_DP}
\end{equation} 
with $\varrho(\vr)$ the (supposedly isotropic) mass density of the particle. By introducing the Fourier transform $\tilde{\varrho}(\vk)$ of the latter, one arrives at the form (\ref{eq:ME}) with $g(k) = \left| \tilde{\varrho} (\vk) \right|^2/m^2$.

In the CSL model the function $g(k)/k^2$ is assumed to be a Gaussian whose inverse width is set to about 100\,nm \cite{Ghirardi1990b,Bassi2003,Vacchini2007b}. Extensive studies on whether such collapse models can be probed in mechanical superposition experiments \cite{Bassi2005,Adler2009,Romero-Isart2011b,Nimmrichter2011a,Kaltenbaek2012,Bassi2012,Bahrami2014,Bateman2014,Nimmrichter2014,Bahrami2014b} may also serve, via the common master equation (\ref{eq:ME}), as a test criterion for stochastic SN equations. 

In general, $g(k)$ can be any positive and integrable function. 
The above form (\ref{eq:ME}) then falls into a generic class of Galileian-covariant 
master equations \footnote{In the most general case of a (norm-bounded) Galileian-covariant master equation, the momentum kick operators in (\ref{eq:ME}) are replaced by Weyl unitaries representing shifts in both position and momentum. These operators are then distributed according to a positive and integrable function $g(r,k)$, see \cite{Holevo1993,Nimmrichter2013}.}, 
which gradually ``classicalize'' the state of motion $\rho$, rendering it indistinguishable from a classical mixture in phase space \cite{Holevo1993,Nimmrichter2013}. In fact, Eq.~(\ref{eq:ME}) resembles collisional decoherence-type master equations \cite{Vacchini2009}, which describe the decay of spatial coherence in combination with momentum diffusion, and whose stable pointer-state solutions are solitonic wave packets moving on Newtonian trajectories \cite{Busse2009,Busse2010a}. 

\subsection{Discussion}

Equations (\ref{eq:SSE_PDP}) and (\ref{eq:SSE_W}) demonstrate that there exist mathematically simple stochastic extensions of the SN equation, which cancel the nonlinearity (\ref{eq:HSN}) in the statistically averaged time evolution (\ref{eq:ME}). 
For this, the originally unbounded SN potential (\ref{eq:VSN}) is to be regularized; and assuming the associated function $g(k)$ vanishes beyond a characteristic width $\sigma_k$, the jumps (or the noise amplitude) can be neglected in the stochastic SN equation for $|\vk|\gtrsim \sigma_k$. 

It must be stressed that the regularizing function $g(k)$ may not be easily dispensed with. Omitting it in (\ref{eq:HSN}) and (\ref{eq:ME}) would, for instance, result in a divergent average momentum diffusion rate $\partial_t \la \ovp^2\ra$.
One may even argue that a 'high-energy cutoff' in the form of $g(k)$ should anyhow appear in the SN potential (\ref{eq:HSN}), if the latter is supposed to be the effective low-energy remnant of quantized gravity. 

Other, more complicated stochastic extensions can be conceived as well, e.g.~via unitary mixing of the jump operators (\ref{eq:A}). 
A straightforward example can be constructed by Fourier transforming the jump operators,
\begin{equation}
 \oB_{\psi} (\vs) = \int \diff^3 k \frac{\sqrt{g(k)}}{(2\pi)^{3/2} k} e^{i\vk\cdot\vs} \oA_{\psi} (\vk) = \int \diff^3 k \frac{\sqrt{g(k)}}{(2\pi)^{3/2} k} \left[ e^{i\vk\cdot(\vs-\ovr)} + i \la \psi | e^{i\vk\cdot(\vs-\ovr)} |\psi\ra \right]. \label{eq:B}
\end{equation}
The associated piecewise deterministic extension,
\begin{eqnarray}
 |\diff \psi \ra &=& -\frac{i}{\hbar} \left( \oH + \oH^\SN_{\psi} \right) |\psi\ra \diff t + \int\diff^3 s \left( \frac{\oB_{\psi} (\vs)}{\left\| \oB_{\psi} (\vs) |\psi\ra \right\|} - 1 \right) |\psi\ra \diff N_{\vs}, \label{eq:SSE_PDP_B}
\end{eqnarray} 
leads again to the same statistically averaged master equation (\ref{eq:ME}) as before, if $\erw{\diff N_{\vs}} / \diff t = Gm^2 \left\| \oB_{\psi} (\vs) |\psi\ra \right\|^2/2\pi^2\hbar$; the diffusive form follows by analogy.

The physical meaning of the presented stochastic Schr\"{o}dinger equations remains to be clarified, not least the peculiar form of the jump operators (\ref{eq:A}) or (\ref{eq:B}). 
They are given in terms of momentum kick operators to which an expectation value is added with a phase. 
This construction, in combination with the regularizing function $g(k)$, is the price to pay for reconciling the nonlinear SN equation for the state vector $|\psi\ra$ with the linear decoherence-type master equation (\ref{eq:ME}) for the ensemble state $\rho$. It is a price hard to bargain if nonlinear time evolutions of $\rho$ and possible violations of no-signalling are to be avoided. 

\section{Many-particle generalization} \label{sec:manybody}

As already mentioned, the single-particle SN equation can be put into question because of the mean-field origin of the SN potential \cite{Anastopoulos2014}. However, many-particle formulations of the SN equation exist both for the center-of-mass motion of harmonically bound compounds \cite{Yang2013}, and for the general case of a dynamical $N$-body system with total mass $M=m_1 + m_2 + \ldots + m_N$ \cite{Diosi1984,Giulini2014}. There, an $N$-particle wave function $\Psi \left( \vr_1,\ldots,\vr_N \right)$ is subject to the total gravitational potential 
\begin{equation}
 V^\SN_{\Psi} (\vr_1,\ldots,\vr_N) = -G \sum_{n,\ell=1}^N m_n m_\ell \int \diff^3 r'_1 \ldots \diff^3 r'_N \frac{\left| \Psi \left( \vr'_1,\ldots,\vr'_N \right) \right|^2}{\left| \vr_n - \vr'_\ell \right|}, \label{eq:VSN_N}
\end{equation} 
consisting of both mutual interactions and self-interactions.

The stochastic extensions given in Sect.~\ref{sec:extensions} are readily generalized to $N$-particle systems of distinguishable or indistinguishable species: We consistently replace all unitary single-particle momentum shift operators $\exp(-i\vk\cdot\ovr)$ by non-unitary, mass-weighted sums of single-particle shifts,
\begin{equation}
 \oM_{\vk} = \sum_{n=1}^N \frac{m_n}{M} e^{-i\vk\cdot\ovr_n}. \label{eq:M_N}
\end{equation}
The $N$-particle SN Hamiltonian can now be expressed in terms of these operators, after applying the same Fourier transformation and regularization procedure as for the single-particle case (\ref{eq:HSN}),
\begin{equation}
 \oH^\SN_{\Psi} = V^\SN_{\Psi} (\ovr_1,\ldots,\ovr_N) = - \int\diff^3 k \, \frac{GM^2}{2\pi^2 k^2} g(k) \la \Psi | \oM^\da_{\vk} |\Psi\ra  \oM_{\vk}. \label{eq:HSN_N}
\end{equation} 
The same replacement rule applies to the nonlinear jump operators (\ref{eq:A}) as well,
\begin{equation}
 \oA_{\Psi} (\vk) = \oM_{\vk} + i \la \Psi | \oM_{\vk} |\Psi\ra. \label{eq:A_N}
\end{equation} 
The piecewise deterministic extension (\ref{eq:SSE_PDP}) then generalizes to
\begin{eqnarray}
 |\diff \Psi \ra &=& -\frac{i}{\hbar} \left( \oH + \oH^\SN_{\Psi} \right) |\Psi\ra \diff t + \int\diff^3 k \left[ \frac{\oA_{\Psi} (\vk)}{\left\| \oA_{\Psi} (\vk) |\Psi\ra \right\|} - 1 \right] |\Psi\ra \diff N_{\vk} \nonumber \\
 && + \frac{1}{2} \int\diff^3 k \frac{GM^2}{2\pi^2  \hbar k^2} g(k) \left[ \left\| \oA_{\Psi} (\vk) |\Psi\ra \right\|^2 - \oA^\da_{\Psi} (\vk) \oA_{\Psi} (\vk) \right] |\Psi\ra \diff t, \label{eq:SSE_PDP_N}
\end{eqnarray} 
with $\erw{\diff N_{\vk}} = (GM^2/2\pi^2 \hbar k^2) g(k) \left\|\oA_{\Psi} (\vk) |\Psi\ra \right\|^2 \diff t$. 
Note that the last term is required here for norm conservation in the statistical average, because the operators (\ref{eq:M_N}) are non-unitary. This additional term vanishes only in the single-particle case.

The diffusive extension (\ref{eq:SSE_PDP}) generalizes to
\begin{eqnarray}
 |\diff \Psi \ra &=& -\frac{i}{\hbar} \left( \oH + \oH^\SN_{\Psi} \right) |\Psi\ra \diff t + \int\diff^3 k \, \oA_{\Psi} (\vk) |\Psi\ra \diff W_{\vk} \nonumber \\
 && - \frac{1}{2} \int\diff^3 k \frac{GM^2}{2\pi^2 \hbar k^2} g(k) \oA^\da_{\Psi} (\vk) \oA_{\Psi} (\vk) |\Psi\ra \diff t, \label{eq:SSE_W_N}
\end{eqnarray} 
with the Wiener increments fulfilling $\erw{\diff W^{*}_{\vk} \diff W_{\vq}} = \delta(\vk-\vq) (GM^2/2\pi^2 \hbar k^2) g(k) \diff t$ in It\^{o} calculus.

Both cases yield the same master equation for the statistically averaged time evolution of the density operator,
\begin{equation}
 \partial_t \rho = -\frac{i}{\hbar} \left[ \oH,\rho \right] + \int\diff^3 k \, \frac{GM^2 g(k)}{2\pi^2 \hbar k^2} \left( \oM_{\vk}\rho \oM^\da_{\vk} -\frac{1}{2} \left\{\oM^\da_{\vk}\oM_{\vk},\rho\right\} \right). \label{eq:ME_N}
\end{equation} 
This is once again confirmed by computing the expectation value $\erw{\diff\left( |\psi\ra \la \psi| \right)}/\diff t$ and noting that $\oM^\da_{\vk} = \oM_{-\vk}$.
It is important to notice that this $N$-particle master equation still falls under the class of generic ``classicalizing'' modifications of the von Neumann equation, which are invariant under Galileian symmetry transformations \cite{Nimmrichter2013}. Moreover, it still resembles the CSL model if $g(k)/k^2$ is chosen to be a Gaussian \cite{Bassi2003}. 
It may also serve as an $N$-body version of Di\'{o}si's master equation (\ref{eq:ME_DP}), which assumes that the mass of every particle is distributed according to the same (real and isotropic) distribution function $f(\vr)$, $\int\diff^3 r\, f(\vr) = 1$. The mass density of each particle is then given by $\varrho_n (\vr) = m_n f(\vr)$. Setting $g(k) = |\tilde{f}(\vk)|^2$, the above master equation (\ref{eq:ME_N}) can be rewritten as \cite{Ghirardi1990a}
\begin{equation}
 \partial_t \rho = -\frac{i}{\hbar} \left[ \oH,\rho \right] - \frac{G}{2\hbar} \sum_{n,\ell=1}^N \int \frac{\diff^3 s_1 \diff^3 s_2}{|\vs_1 - \vs_2|} \left[ \varrho_n \left( \vs_1 - \ovr_n \right), \left[ \varrho_{\ell} \left( \vs_2 - \ovr_{\ell} \right), \rho \right] \right]. \label{eq:ME_DP_N}
\end{equation} 
This is a generalization of the single-particle DP model (\ref{eq:ME_DP}) describing mutual gravity and self-gravity in an equal manner. The fact that the unitary part of Eqs.~(\ref{eq:ME_N}) and (\ref{eq:ME_DP_N}) does not involve the standard gravitational pair interaction, raises the question whether this equal treatment of self and mutual gravity in the many-body description is meaningful.

\section{Conclusion}

We presented two stochastic versions of the SN equation for self-gravitating quantum particles, which circumvent the violation of no-signalling by regularizing the SN potential and compensating it with a random jump or diffusion process. This renders the ensemble-averaged time evolution (\ref{eq:ME}) linear. Both single-particle equations (\ref{eq:SSE_PDP}) and (\ref{eq:SSE_W}) can be generalized consistently to many-particle systems, Eqs.~(\ref{eq:SSE_PDP_N}) and (\ref{eq:SSE_W_N}), which results in the linear master equation (\ref{eq:ME_N}). 
The latter serves as a link between the many-particle formulations of the original SN equation \cite{Diosi1984,Yang2013,Giulini2014} and the many-body versions of the CSL model \cite{Bassi2003}, of the DP model \cite{Ghirardi1990a}, 
and of Galileian-covariant ``classicalizing'' modifications of the von Neumann equation in general \cite{Nimmrichter2013}. 

A common feature of the presented stochastic equations is the peculiar form of the jump or noise operators (\ref{eq:A}) and (\ref{eq:A_N}). They split a wave function into a momentum-shifted and an unshifted part, where the relative weight depends on the initial delocalization in momentum space. 
An interesting direction for further study would be to analyze the quantum trajectories in the presence of such random jumps (or Brownian noise). 
Might these jumps be signatures of a deeper theory of quantum gravity \cite{Smolin2001a}? 

\begin{acknowledgments}
We thank A.~Bassi for helpful discussions, and 
we acknowledge support by the European Commission within NANOQUESTFIT (No. 304886).
\end{acknowledgments}


\begin{thebibliography}{40}%
\makeatletter
\providecommand \@ifxundefined [1]{%
 \@ifx{#1\undefined}
}%
\providecommand \@ifnum [1]{%
 \ifnum #1\expandafter \@firstoftwo
 \else \expandafter \@secondoftwo
 \fi
}%
\providecommand \@ifx [1]{%
 \ifx #1\expandafter \@firstoftwo
 \else \expandafter \@secondoftwo
 \fi
}%
\providecommand \natexlab [1]{#1}%
\providecommand \enquote  [1]{``#1''}%
\providecommand \bibnamefont  [1]{#1}%
\providecommand \bibfnamefont [1]{#1}%
\providecommand \citenamefont [1]{#1}%
\providecommand \href@noop [0]{\@secondoftwo}%
\providecommand \href [0]{\begingroup \@sanitize@url \@href}%
\providecommand \@href[1]{\@@startlink{#1}\@@href}%
\providecommand \@@href[1]{\endgroup#1\@@endlink}%
\providecommand \@sanitize@url [0]{\catcode `\\12\catcode `\$12\catcode
  `\&12\catcode `\#12\catcode `\^12\catcode `\_12\catcode `\%12\relax}%
\providecommand \@@startlink[1]{}%
\providecommand \@@endlink[0]{}%
\providecommand \url  [0]{\begingroup\@sanitize@url \@url }%
\providecommand \@url [1]{\endgroup\@href {#1}{\urlprefix }}%
\providecommand \urlprefix  [0]{URL }%
\providecommand \Eprint [0]{\href }%
\providecommand \doibase [0]{http://dx.doi.org/}%
\providecommand \selectlanguage [0]{\@gobble}%
\providecommand \bibinfo  [0]{\@secondoftwo}%
\providecommand \bibfield  [0]{\@secondoftwo}%
\providecommand \translation [1]{[#1]}%
\providecommand \BibitemOpen [0]{}%
\providecommand \bibitemStop [0]{}%
\providecommand \bibitemNoStop [0]{.\EOS\space}%
\providecommand \EOS [0]{\spacefactor3000\relax}%
\providecommand \BibitemShut  [1]{\csname bibitem#1\endcsname}%
\let\auto@bib@innerbib\@empty
\bibitem [{\citenamefont {Ruffini}\ and\ \citenamefont
  {Bonazolla}(1969)}]{Ruffini1969}%
  \BibitemOpen
  \bibfield  {author} {\bibinfo {author} {\bibfnamefont {R.}~\bibnamefont
  {Ruffini}}\ and\ \bibinfo {author} {\bibfnamefont {S.}~\bibnamefont
  {Bonazolla}},\ }\href {\doibase 10.1103/PhysRev.187.1767} {\bibfield
  {journal} {\bibinfo  {journal} {Phys. Rev.}\ }\textbf {\bibinfo {volume}
  {187}},\ \bibinfo {pages} {1767} (\bibinfo {year} {1969})}\BibitemShut
  {NoStop}%
\bibitem [{\citenamefont {Di\'{o}si}(1984)}]{Diosi1984}%
  \BibitemOpen
  \bibfield  {author} {\bibinfo {author} {\bibfnamefont {L.}~\bibnamefont
  {Di\'{o}si}},\ }\href {\doibase 10.1016/0375-9601(84)90397-9} {\bibfield
  {journal} {\bibinfo  {journal} {Phys. Lett. A}\ }\textbf {\bibinfo {volume}
  {105}},\ \bibinfo {pages} {199} (\bibinfo {year} {1984})}\BibitemShut
  {NoStop}%
\bibitem [{\citenamefont {Carlip}(2008)}]{Carlip2008}%
  \BibitemOpen
  \bibfield  {author} {\bibinfo {author} {\bibfnamefont {S.}~\bibnamefont
  {Carlip}},\ }\href {http://iopscience.iop.org/0264-9381/25/15/154010}
  {\bibfield  {journal} {\bibinfo  {journal} {Class. Quantum Gravity}\ }\textbf
  {\bibinfo {volume} {25}},\ \bibinfo {pages} {154010} (\bibinfo {year}
  {2008})}\BibitemShut {NoStop}%
\bibitem [{\citenamefont {Giulini}\ and\ \citenamefont
  {Gro\ss{}ardt}(2011)}]{Giulini2011}%
  \BibitemOpen
  \bibfield  {author} {\bibinfo {author} {\bibfnamefont {D.}~\bibnamefont
  {Giulini}}\ and\ \bibinfo {author} {\bibfnamefont {A.}~\bibnamefont
  {Gro\ss{}ardt}},\ }\href {\doibase 10.1088/0264-9381/28/19/195026} {\bibfield
  {journal} {\bibinfo  {journal} {Class. Quantum Gravity}\ }\textbf {\bibinfo
  {volume} {28}},\ \bibinfo {pages} {195026} (\bibinfo {year}
  {2011})}\BibitemShut {NoStop}%
\bibitem [{\citenamefont {Giulini}\ and\ \citenamefont
  {Gro\ss{}ardt}(2012)}]{Giulini2012}%
  \BibitemOpen
  \bibfield  {author} {\bibinfo {author} {\bibfnamefont {D.}~\bibnamefont
  {Giulini}}\ and\ \bibinfo {author} {\bibfnamefont {A.}~\bibnamefont
  {Gro\ss{}ardt}},\ }\href {\doibase 10.1088/0264-9381/29/21/215010} {\bibfield
  {journal} {\bibinfo  {journal} {Class. Quantum Gravity}\ }\textbf {\bibinfo
  {volume} {29}},\ \bibinfo {pages} {215010} (\bibinfo {year}
  {2012})}\BibitemShut {NoStop}%
\bibitem [{\citenamefont {Anastopoulos}\ and\ \citenamefont
  {Hu}(2014)}]{Anastopoulos2014}%
  \BibitemOpen
  \bibfield  {author} {\bibinfo {author} {\bibfnamefont {C.}~\bibnamefont
  {Anastopoulos}}\ and\ \bibinfo {author} {\bibfnamefont {B.~L.}\ \bibnamefont
  {Hu}},\ }\href {\doibase 10.1088/1367-2630/16/8/085007} {\bibfield  {journal}
  {\bibinfo  {journal} {New J. Phys.}\ }\textbf {\bibinfo {volume} {16}},\
  \bibinfo {pages} {085007} (\bibinfo {year} {2014})}\BibitemShut {NoStop}%
\bibitem [{\citenamefont {Giulini}\ and\ \citenamefont
  {Gro\ss{}ardt}(2014)}]{Giulini2014}%
  \BibitemOpen
  \bibfield  {author} {\bibinfo {author} {\bibfnamefont {D.}~\bibnamefont
  {Giulini}}\ and\ \bibinfo {author} {\bibfnamefont {A.}~\bibnamefont
  {Gro\ss{}ardt}},\ }\href {\doibase 10.1088/1367-2630/16/7/075005} {\bibfield
  {journal} {\bibinfo  {journal} {New J. Phys.}\ }\textbf {\bibinfo {volume}
  {16}},\ \bibinfo {pages} {075005} (\bibinfo {year} {2014})}\BibitemShut
  {NoStop}%
\bibitem [{\citenamefont {Gisin}(1989)}]{Gisin1989}%
  \BibitemOpen
  \bibfield  {author} {\bibinfo {author} {\bibfnamefont {N.}~\bibnamefont
  {Gisin}},\ }\href
  {http://inis.iaea.org/Search/search.aspx?orig\_q=RN:20077415} {\bibfield
  {journal} {\bibinfo  {journal} {Helv. Phys. Act.}\ }\textbf {\bibinfo
  {volume} {62}},\ \bibinfo {pages} {363} (\bibinfo {year} {1989})}\BibitemShut
  {NoStop}%
\bibitem [{\citenamefont {Bassi}\ and\ \citenamefont
  {Ghirardi}(2003)}]{Bassi2003}%
  \BibitemOpen
  \bibfield  {author} {\bibinfo {author} {\bibfnamefont {A.}~\bibnamefont
  {Bassi}}\ and\ \bibinfo {author} {\bibfnamefont {G.}~\bibnamefont
  {Ghirardi}},\ }\href@noop {} {\bibfield  {journal} {\bibinfo  {journal}
  {Phys. Rep.}\ }\textbf {\bibinfo {volume} {379}},\ \bibinfo {pages} {257}
  (\bibinfo {year} {2003})}\BibitemShut {NoStop}%
\bibitem [{\citenamefont {Bahrami}\ \emph
  {et~al.}(2014{\natexlab{a}})\citenamefont {Bahrami}, \citenamefont
  {Gro\ss{}ardt}, \citenamefont {Donadi},\ and\ \citenamefont
  {Bassi}}]{Bahrami2014a}%
  \BibitemOpen
  \bibfield  {author} {\bibinfo {author} {\bibfnamefont {M.}~\bibnamefont
  {Bahrami}}, \bibinfo {author} {\bibfnamefont {A.}~\bibnamefont
  {Gro\ss{}ardt}}, \bibinfo {author} {\bibfnamefont {S.}~\bibnamefont {Donadi}},
  \ and\ \bibinfo {author} {\bibfnamefont {A.}~\bibnamefont {Bassi}},\ }\href
  {http://arxiv.org/abs/1407.4370} {(\bibinfo {year}
  {2014}{\natexlab{a}})},\ \Eprint {http://arxiv.org/abs/1407.4370}
  {arXiv:1407.4370} \BibitemShut {NoStop}%
\bibitem [{\citenamefont {Karolyhazy}(1966)}]{Karolyhazy1966}%
  \BibitemOpen
  \bibfield  {author} {\bibinfo {author} {\bibfnamefont {F.}~\bibnamefont
  {Karolyhazy}},\ }\href {\doibase 10.1007/BF02717926} {\bibfield  {journal}
  {\bibinfo  {journal} {Nuovo Cim. A}\ }\textbf {\bibinfo {volume} {42}},\
  \bibinfo {pages} {390} (\bibinfo {year} {1966})}\BibitemShut {NoStop}%
\bibitem [{\citenamefont {Ghirardi}\ \emph {et~al.}(1986)\citenamefont
  {Ghirardi}, \citenamefont {Rimini},\ and\ \citenamefont
  {Weber}}]{Ghirardi1986}%
  \BibitemOpen
  \bibfield  {author} {\bibinfo {author} {\bibfnamefont {G.~C.}\ \bibnamefont
  {Ghirardi}}, \bibinfo {author} {\bibfnamefont {A.}~\bibnamefont {Rimini}}, \
  and\ \bibinfo {author} {\bibfnamefont {T.}~\bibnamefont {Weber}},\
  }\href@noop {} {\bibfield  {journal} {\bibinfo  {journal} {Phys. Rev. D}\
  }\textbf {\bibinfo {volume} {34}},\ \bibinfo {pages} {470} (\bibinfo {year}
  {1986})}\BibitemShut {NoStop}%
\bibitem [{\citenamefont {Di\'{o}si}(1987)}]{Diosi1987}%
  \BibitemOpen
  \bibfield  {author} {\bibinfo {author} {\bibfnamefont {L.}~\bibnamefont
  {Di\'{o}si}},\ }\href {\doibase 10.1016/0375-9601(87)90681-5} {\bibfield
  {journal} {\bibinfo  {journal} {Phys. Lett. A}\ }\textbf {\bibinfo {volume}
  {120}},\ \bibinfo {pages} {377} (\bibinfo {year} {1987})}\BibitemShut
  {NoStop}%
\bibitem [{\citenamefont {Ghirardi}\ \emph
  {et~al.}(1990{\natexlab{a}})\citenamefont {Ghirardi}, \citenamefont
  {Pearle},\ and\ \citenamefont {Rimini}}]{Ghirardi1990b}%
  \BibitemOpen
  \bibfield  {author} {\bibinfo {author} {\bibfnamefont {G.~C.}\ \bibnamefont
  {Ghirardi}}, \bibinfo {author} {\bibfnamefont {P.}~\bibnamefont {Pearle}}, \
  and\ \bibinfo {author} {\bibfnamefont {A.}~\bibnamefont {Rimini}},\
  }\href@noop {} {\bibfield  {journal} {\bibinfo  {journal} {Phys. Rev. A}\
  }\textbf {\bibinfo {volume} {42}},\ \bibinfo {pages} {78} (\bibinfo {year}
  {1990}{\natexlab{a}})}\BibitemShut {NoStop}%
\bibitem [{\citenamefont {Penrose}(1996)}]{Penrose1996}%
  \BibitemOpen
  \bibfield  {author} {\bibinfo {author} {\bibfnamefont {R.}~\bibnamefont
  {Penrose}},\ }\href {http://www.springerlink.com/index/k75046wh3668l654.pdf}
  {\bibfield  {journal} {\bibinfo  {journal} {Gen. Relativ. Gravit.}\ }\textbf
  {\bibinfo {volume} {28}},\ \bibinfo {pages} {581} (\bibinfo {year}
  {1996})}\BibitemShut {NoStop}%
\bibitem [{\citenamefont {Bassi}\ \emph {et~al.}(2013)\citenamefont {Bassi},
  \citenamefont {Lochan}, \citenamefont {Satin}, \citenamefont {Singh},\ and\
  \citenamefont {Ulbricht}}]{Bassi2012}%
  \BibitemOpen
  \bibfield  {author} {\bibinfo {author} {\bibfnamefont {A.}~\bibnamefont
  {Bassi}}, \bibinfo {author} {\bibfnamefont {K.}~\bibnamefont {Lochan}},
  \bibinfo {author} {\bibfnamefont {S.}~\bibnamefont {Satin}}, \bibinfo
  {author} {\bibfnamefont {T.~P.}\ \bibnamefont {Singh}}, \ and\ \bibinfo
  {author} {\bibfnamefont {H.}~\bibnamefont {Ulbricht}},\ }\href {\doibase
  10.1103/RevModPhys.85.471} {\bibfield  {journal} {\bibinfo  {journal} {Rev.
  Mod. Phys.}\ }\textbf {\bibinfo {volume} {85}},\ \bibinfo {pages} {471}
  (\bibinfo {year} {2013})}\BibitemShut {NoStop}%
\bibitem [{\citenamefont {Adler}(2004)}]{Adler2004}%
  \BibitemOpen
  \bibfield  {author} {\bibinfo {author} {\bibfnamefont {S.~L.}\ \bibnamefont
  {Adler}},\ }\href
  {http://www.amazon.com/Quantum-Theory-Emergent-Phenomenon-Statistical/dp/0521831946}
  {\emph {\bibinfo {title} {{Quantum Theory as an Emergent Phenomenon: The
  Statistical Mechanics of Matrix Models as the Precursor of Quantum Field
  Theory}}}}\ (\bibinfo  {publisher} {Cambridge University Press},\ \bibinfo
  {year} {2004})\BibitemShut {NoStop}%
\bibitem [{\citenamefont {Vacchini}(2007)}]{Vacchini2007b}%
  \BibitemOpen
  \bibfield  {author} {\bibinfo {author} {\bibfnamefont {B.}~\bibnamefont
  {Vacchini}},\ }\href@noop {} {\bibfield  {journal} {\bibinfo  {journal} {J.
  Phys. A Math. Theor.}\ }\textbf {\bibinfo {volume} {40}},\ \bibinfo {pages}
  {2463} (\bibinfo {year} {2007})}\BibitemShut {NoStop}%
\bibitem [{\citenamefont {Vacchini}\ and\ \citenamefont
  {Hornberger}(2009)}]{Vacchini2009}%
  \BibitemOpen
  \bibfield  {author} {\bibinfo {author} {\bibfnamefont {B.}~\bibnamefont
  {Vacchini}}\ and\ \bibinfo {author} {\bibfnamefont {K.}~\bibnamefont
  {Hornberger}},\ }\href {\doibase 10.1016/j.physrep.2009.06.001} {\bibfield
  {journal} {\bibinfo  {journal} {Phys. Rep.}\ }\textbf {\bibinfo {volume}
  {478}},\ \bibinfo {pages} {71} (\bibinfo {year} {2009})}\BibitemShut
  {NoStop}%
\bibitem [{\citenamefont {Rosenfeld}(1963)}]{Rosenfeld1963}%
  \BibitemOpen
  \bibfield  {author} {\bibinfo {author} {\bibfnamefont {L.}~\bibnamefont
  {Rosenfeld}},\ }\href {\doibase 10.1016/0029-5582(63)90279-7} {\bibfield
  {journal} {\bibinfo  {journal} {Nucl. Phys.}\ }\textbf {\bibinfo {volume}
  {40}},\ \bibinfo {pages} {353} (\bibinfo {year} {1963})}\BibitemShut
  {NoStop}%
\bibitem [{\citenamefont {Breuer}\ and\ \citenamefont
  {Petruccione}(2002)}]{Breuer2002}%
  \BibitemOpen
  \bibfield  {author} {\bibinfo {author} {\bibfnamefont {H.-P.}\ \bibnamefont
  {Breuer}}\ and\ \bibinfo {author} {\bibfnamefont {F.}~\bibnamefont
  {Petruccione}},\ }\href
  {http://books.google.com/books?id=0Yx5VzaMYm8C\&pgis=1} {\emph {\bibinfo
  {title} {{The Theory of Open Quantum Systems}}}}\ (\bibinfo  {publisher}
  {Oxford University Press},\ \bibinfo {year} {2002})\ p.\ \bibinfo {pages}
  {625}\BibitemShut {NoStop}%
\bibitem [{\citenamefont {Wiseman}\ and\ \citenamefont
  {Milburn}(2010)}]{Wiseman2010}%
  \BibitemOpen
  \bibfield  {author} {\bibinfo {author} {\bibfnamefont {H.~M.}\ \bibnamefont
  {Wiseman}}\ and\ \bibinfo {author} {\bibfnamefont {G.~J.}\ \bibnamefont
  {Milburn}},\ }\href@noop {} {\emph {\bibinfo {title} {{Quantum measurement
  and control}}}}\ (\bibinfo  {publisher} {Cambridge University Press},\
  \bibinfo {year} {2010})\BibitemShut {NoStop}%
\bibitem [{\citenamefont {Di\'{o}si}(1989)}]{Diosi1989}%
  \BibitemOpen
  \bibfield  {author} {\bibinfo {author} {\bibfnamefont {L.}~\bibnamefont
  {Di\'{o}si}},\ }\href {\doibase 10.1103/PhysRevA.40.1165} {\bibfield
  {journal} {\bibinfo  {journal} {Phys. Rev. A}\ }\textbf {\bibinfo {volume}
  {40}},\ \bibinfo {pages} {1165} (\bibinfo {year} {1989})}\BibitemShut
  {NoStop}%
\bibitem [{\citenamefont {Bahrami}\ \emph
  {et~al.}(2014{\natexlab{b}})\citenamefont {Bahrami}, \citenamefont {Smirne},\
  and\ \citenamefont {Bassi}}]{Bahrami2014b}%
  \BibitemOpen
  \bibfield  {author} {\bibinfo {author} {\bibfnamefont {M.}~\bibnamefont
  {Bahrami}}, \bibinfo {author} {\bibfnamefont {A.}~\bibnamefont {Smirne}}, \
  and\ \bibinfo {author} {\bibfnamefont {A.}~\bibnamefont {Bassi}},\ }\href
  {http://arxiv.org/abs/1408.6460} {\  (\bibinfo {year}
  {2014}{\natexlab{b}})},\ \Eprint {http://arxiv.org/abs/1408.6460}
  {arXiv:1408.6460} \BibitemShut {NoStop}%
\bibitem [{\citenamefont {Bassi}\ \emph {et~al.}(2005)\citenamefont {Bassi},
  \citenamefont {Ippoliti},\ and\ \citenamefont {Adler}}]{Bassi2005}%
  \BibitemOpen
  \bibfield  {author} {\bibinfo {author} {\bibfnamefont {A.}~\bibnamefont
  {Bassi}}, \bibinfo {author} {\bibfnamefont {E.}~\bibnamefont {Ippoliti}}, \
  and\ \bibinfo {author} {\bibfnamefont {S.}~\bibnamefont {Adler}},\
  }\href@noop {} {\bibfield  {journal} {\bibinfo  {journal} {Phys. Rev. Lett.}\
  }\textbf {\bibinfo {volume} {94}},\ \bibinfo {pages} {030401} (\bibinfo {year}
  {2005})}\BibitemShut {NoStop}%
\bibitem [{\citenamefont {Adler}\ and\ \citenamefont
  {Bassi}(2009)}]{Adler2009}%
  \BibitemOpen
  \bibfield  {author} {\bibinfo {author} {\bibfnamefont {S.~L.}\ \bibnamefont
  {Adler}}\ and\ \bibinfo {author} {\bibfnamefont {A.}~\bibnamefont {Bassi}},\
  }\href@noop {} {\bibfield  {journal} {\bibinfo  {journal} {Science (80-. ).}\
  }\textbf {\bibinfo {volume} {325}},\ \bibinfo {pages} {275} (\bibinfo {year}
  {2009})}\BibitemShut {NoStop}%
\bibitem [{\citenamefont {Romero-Isart}\ \emph {et~al.}(2011)\citenamefont
  {Romero-Isart}, \citenamefont {Pflanzer}, \citenamefont {Blaser},
  \citenamefont {Kaltenbaek}, \citenamefont {Kiesel}, \citenamefont
  {Aspelmeyer},\ and\ \citenamefont {Cirac}}]{Romero-Isart2011b}%
  \BibitemOpen
  \bibfield  {author} {\bibinfo {author} {\bibfnamefont {O.}~\bibnamefont
  {Romero-Isart}}, \bibinfo {author} {\bibfnamefont {A.~C.}\ \bibnamefont
  {Pflanzer}}, \bibinfo {author} {\bibfnamefont {F.}~\bibnamefont {Blaser}},
  \bibinfo {author} {\bibfnamefont {R.}~\bibnamefont {Kaltenbaek}}, \bibinfo
  {author} {\bibfnamefont {N.}~\bibnamefont {Kiesel}}, \bibinfo {author}
  {\bibfnamefont {M.}~\bibnamefont {Aspelmeyer}}, \ and\ \bibinfo {author}
  {\bibfnamefont {J.~I.}\ \bibnamefont {Cirac}},\ }\href@noop {} {\bibfield
  {journal} {\bibinfo  {journal} {Phys. Rev. Lett.}\ }\textbf {\bibinfo
  {volume} {107}},\ \bibinfo {pages} {020405} (\bibinfo {year}
  {2011})}\BibitemShut {NoStop}%
\bibitem [{\citenamefont {Nimmrichter}\ \emph {et~al.}(2011)\citenamefont
  {Nimmrichter}, \citenamefont {Hornberger}, \citenamefont {Haslinger},\ and\
  \citenamefont {Arndt}}]{Nimmrichter2011a}%
  \BibitemOpen
  \bibfield  {author} {\bibinfo {author} {\bibfnamefont {S.}~\bibnamefont
  {Nimmrichter}}, \bibinfo {author} {\bibfnamefont {K.}~\bibnamefont
  {Hornberger}}, \bibinfo {author} {\bibfnamefont {P.}~\bibnamefont
  {Haslinger}}, \ and\ \bibinfo {author} {\bibfnamefont {M.}~\bibnamefont
  {Arndt}},\ }\href@noop {} {\bibfield  {journal} {\bibinfo  {journal} {Phys.
  Rev. A}\ }\textbf {\bibinfo {volume} {83}},\ \bibinfo {pages} {043621}
  (\bibinfo {year} {2011})}\BibitemShut {NoStop}%
\bibitem [{\citenamefont {Kaltenbaek}\ \emph {et~al.}(2012)\citenamefont
  {Kaltenbaek}, \citenamefont {Hechenblaikner}, \citenamefont {Kiesel},
  \citenamefont {Romero-Isart}, \citenamefont {Schwab}, \citenamefont
  {Johann},\ and\ \citenamefont {Aspelmeyer}}]{Kaltenbaek2012}%
  \BibitemOpen
  \bibfield  {author} {\bibinfo {author} {\bibfnamefont {R.}~\bibnamefont
  {Kaltenbaek}}, \bibinfo {author} {\bibfnamefont {G.}~\bibnamefont
  {Hechenblaikner}}, \bibinfo {author} {\bibfnamefont {N.}~\bibnamefont
  {Kiesel}}, \bibinfo {author} {\bibfnamefont {O.}~\bibnamefont
  {Romero-Isart}}, \bibinfo {author} {\bibfnamefont {K.~C.}\ \bibnamefont
  {Schwab}}, \bibinfo {author} {\bibfnamefont {U.}~\bibnamefont {Johann}}, \
  and\ \bibinfo {author} {\bibfnamefont {M.}~\bibnamefont {Aspelmeyer}},\
  }\href {\doibase 10.1007/s10686-012-9292-3} {\bibfield  {journal} {\bibinfo
  {journal} {Exp. Astron.}\ }\textbf {\bibinfo {volume} {34}},\ \bibinfo
  {pages} {123} (\bibinfo {year} {2012})}\BibitemShut {NoStop}%
\bibitem [{\citenamefont {Bahrami}\ \emph
  {et~al.}(2014{\natexlab{c}})\citenamefont {Bahrami}, \citenamefont
  {Paternostro}, \citenamefont {Bassi},\ and\ \citenamefont
  {Ulbricht}}]{Bahrami2014}%
  \BibitemOpen
  \bibfield  {author} {\bibinfo {author} {\bibfnamefont {M.}~\bibnamefont
  {Bahrami}}, \bibinfo {author} {\bibfnamefont {M.}~\bibnamefont
  {Paternostro}}, \bibinfo {author} {\bibfnamefont {A.}~\bibnamefont {Bassi}},
  \ and\ \bibinfo {author} {\bibfnamefont {H.}~\bibnamefont {Ulbricht}},\
  }\href {\doibase 10.1103/PhysRevLett.112.210404} {\bibfield  {journal}
  {\bibinfo  {journal} {Phys. Rev. Lett.}\ }\textbf {\bibinfo {volume} {112}},\
  \bibinfo {pages} {210404} (\bibinfo {year} {2014}{\natexlab{c}})}\BibitemShut
  {NoStop}%
\bibitem [{\citenamefont {Bateman}\ \emph {et~al.}(2014)\citenamefont
  {Bateman}, \citenamefont {Nimmrichter}, \citenamefont {Hornberger},\ and\
  \citenamefont {Ulbricht}}]{Bateman2014}%
  \BibitemOpen
  \bibfield  {author} {\bibinfo {author} {\bibfnamefont {J.}~\bibnamefont
  {Bateman}}, \bibinfo {author} {\bibfnamefont {S.}~\bibnamefont
  {Nimmrichter}}, \bibinfo {author} {\bibfnamefont {K.}~\bibnamefont
  {Hornberger}}, \ and\ \bibinfo {author} {\bibfnamefont {H.}~\bibnamefont
  {Ulbricht}},\ }\href {\doibase 10.1038/ncomms5788} {\bibfield  {journal}
  {\bibinfo  {journal} {Nat. Commun.}\ }\textbf {\bibinfo {volume} {5}},\
  \bibinfo {pages} {4788} (\bibinfo {year} {2014})}\BibitemShut {NoStop}%
\bibitem [{\citenamefont {Nimmrichter}\ \emph {et~al.}(2014)\citenamefont
  {Nimmrichter}, \citenamefont {Hornberger},\ and\ \citenamefont
  {Hammerer}}]{Nimmrichter2014}%
  \BibitemOpen
  \bibfield  {author} {\bibinfo {author} {\bibfnamefont {S.}~\bibnamefont
  {Nimmrichter}}, \bibinfo {author} {\bibfnamefont {K.}~\bibnamefont
  {Hornberger}}, \ and\ \bibinfo {author} {\bibfnamefont {K.}~\bibnamefont
  {Hammerer}},\ }\href {\doibase 10.1103/PhysRevLett.113.020405} {\bibfield
  {journal} {\bibinfo  {journal} {Phys. Rev. Lett.}\ }\textbf {\bibinfo
  {volume} {113}},\ \bibinfo {pages} {020405} (\bibinfo {year}
  {2014})}\BibitemShut {NoStop}%
\bibitem [{Note1()}]{Note1}%
  \BibitemOpen
  \bibinfo {note} {In the most general case of a (norm-bounded)
  Galileian-covariant master equation, the momentum kick operators in (\ref
  {eq:ME}) are replaced by Weyl unitaries representing shifts in both position
  and momentum. These operators are then distributed according to a positive
  and integrable function $g(r,k)$, see \cite
  {Holevo1993,Nimmrichter2013}.}\BibitemShut {Stop}%
\bibitem [{\citenamefont {Holevo}(1993)}]{Holevo1993}%
  \BibitemOpen
  \bibfield  {author} {\bibinfo {author} {\bibfnamefont {A.~S.}\ \bibnamefont
  {Holevo}},\ }\href@noop {} {\bibfield  {journal} {\bibinfo  {journal}
  {Reports Math. Phys.}\ }\textbf {\bibinfo {volume} {32}},\ \bibinfo {pages}
  {211} (\bibinfo {year} {1993})}\BibitemShut {NoStop}%
\bibitem [{\citenamefont {Nimmrichter}\ and\ \citenamefont
  {Hornberger}(2013)}]{Nimmrichter2013}%
  \BibitemOpen
  \bibfield  {author} {\bibinfo {author} {\bibfnamefont {S.}~\bibnamefont
  {Nimmrichter}}\ and\ \bibinfo {author} {\bibfnamefont {K.}~\bibnamefont
  {Hornberger}},\ }\href {\doibase 10.1103/PhysRevLett.110.160403} {\bibfield
  {journal} {\bibinfo  {journal} {Phys. Rev. Lett.}\ }\textbf {\bibinfo
  {volume} {110}},\ \bibinfo {pages} {160403} (\bibinfo {year}
  {2013})}\BibitemShut {NoStop}%
\bibitem [{\citenamefont {Busse}\ and\ \citenamefont
  {Hornberger}(2009)}]{Busse2009}%
  \BibitemOpen
  \bibfield  {author} {\bibinfo {author} {\bibfnamefont {M.}~\bibnamefont
  {Busse}}\ and\ \bibinfo {author} {\bibfnamefont {K.}~\bibnamefont
  {Hornberger}},\ }\href {\doibase 10.1088/1751-8113/42/36/362001} {\bibfield
  {journal} {\bibinfo  {journal} {J. Phys. A Math. Theor.}\ }\textbf {\bibinfo
  {volume} {42}},\ \bibinfo {pages} {362001} (\bibinfo {year}
  {2009})}\BibitemShut {NoStop}%
\bibitem [{\citenamefont {Busse}\ and\ \citenamefont
  {Hornberger}(2010)}]{Busse2010a}%
  \BibitemOpen
  \bibfield  {author} {\bibinfo {author} {\bibfnamefont {M.}~\bibnamefont
  {Busse}}\ and\ \bibinfo {author} {\bibfnamefont {K.}~\bibnamefont
  {Hornberger}},\ }\href@noop {} {\bibfield  {journal} {\bibinfo  {journal} {J.
  Phys. A Math. Theor.}\ }\textbf {\bibinfo {volume} {43}},\ \bibinfo {pages}
  {015303} (\bibinfo {year} {2010})}\BibitemShut {NoStop}%
\bibitem [{\citenamefont {Yang}\ \emph {et~al.}(2012)\citenamefont {Yang},
  \citenamefont {Miao}, \citenamefont {Lee}, \citenamefont {Helou},\ and\
  \citenamefont {Chen}}]{Yang2013}%
  \BibitemOpen
  \bibfield  {author} {\bibinfo {author} {\bibfnamefont {H.}~\bibnamefont
  {Yang}}, \bibinfo {author} {\bibfnamefont {H.}~\bibnamefont {Miao}}, \bibinfo
  {author} {\bibfnamefont {D.-s.}\ \bibnamefont {Lee}}, \bibinfo {author}
  {\bibfnamefont {B.}~\bibnamefont {Helou}}, \ and\ \bibinfo {author}
  {\bibfnamefont {Y.}~\bibnamefont {Chen}},\ }\href {\doibase
  10.1103/PhysRevLett.110.170401} {\bibfield  {journal} {\bibinfo  {journal}
  {Phys. Rev. Lett.}\ }\textbf {\bibinfo {volume} {110}},\ \bibinfo {pages}
  {170401} (\bibinfo {year} {2012})}\BibitemShut {NoStop}%
\bibitem [{\citenamefont {Ghirardi}\ \emph
  {et~al.}(1990{\natexlab{b}})\citenamefont {Ghirardi}, \citenamefont
  {Grassi},\ and\ \citenamefont {Rimini}}]{Ghirardi1990a}%
  \BibitemOpen
  \bibfield  {author} {\bibinfo {author} {\bibfnamefont {G.}~\bibnamefont
  {Ghirardi}}, \bibinfo {author} {\bibfnamefont {R.}~\bibnamefont {Grassi}}, \
  and\ \bibinfo {author} {\bibfnamefont {A.}~\bibnamefont {Rimini}},\ }\href
  {\doibase 10.1103/PhysRevA.42.1057} {\bibfield  {journal} {\bibinfo
  {journal} {Phys. Rev. A}\ }\textbf {\bibinfo {volume} {42}},\ \bibinfo
  {pages} {1057} (\bibinfo {year} {1990}{\natexlab{b}})}\BibitemShut {NoStop}%
\bibitem [{\citenamefont {Smolin}(2001)}]{Smolin2001a}%
  \BibitemOpen
  \bibfield  {author} {\bibinfo {author} {\bibfnamefont {L.}~\bibnamefont
  {Smolin}},\ }\href@noop {} {\emph {\bibinfo {title} {{Three roads to quantum
  gravity}}}}\ (\bibinfo  {publisher} {Basic Books},\ \bibinfo {address} {New
  York},\ \bibinfo {year} {2001})\BibitemShut {NoStop}%
\end{thebibliography}

%

\end{document}